# Agent based Tools for Modeling and Simulation of Self-Organization in Peer-to-Peer, Ad-Hoc and other Complex Networks

Authors: Muaz Niazi and Amir Hussain

Department of Computing Science, University of Stirling, Scotland, UK

## 1. Introduction:

The last decade has seen an unprecedented growth in the scale and complexity of computer networks; networks today are much larger and far more complex than originally anticipated. In turn, the larger and more complex the network, the harder it becomes to test and evaluate its design and eventual implementation. Modeling & simulation (M&S) serves a vital role in the design and development of distributed interacting systems because of their peculiar stochastic nature especially if they involve systems with decentralized self-organizing capabilities [1, 9]. Although networks have grown both in size as well as complexity, the tools to model and simulate them have not scaled at the same rate. This has not gone entirely unnoticed so papers such as [2] have attempted to point out problems with the current simulation methodologies for telecommunications networks such as the use of pseudo-random number generators. However another problem with the existing toolset, which has not received much attention, lies in what initially was their strongest point; their focused orientation on computer networks. Since newer networks have grown in size and complexity in directions such as pervasive computing (including but not limited to Body Area Networks, Vehicular Area Networks and other pervasive networks), the existing M&S tools are not quite designed to deal with these emerging paradigms. With so much unanticipated change in the air, there is a need for alternative flexible tools equipped with effective techniques for modeling modern networks as they evolve into complex systems. This paper seeks to address this important area of research for the M&S community in the domain of computer networks by demonstrating, for the first time, the use of agent-based modeling tools for modeling self-organizing mobile nodes and peer to peer (P2P) networks. We focus on NetLogo, a tool which has proven its worth in the modeling and simulation domain of complex systems. We conclude that not only can agent-based tools model P2P and ad-hoc networks, they can also be used to effectively model complex systems where the network is only a part of the system; and where humans, animals, habitat and other environment-related interactions need to be modeled and simulated as well.

## 2. Background

In the domain of complex networks, being the new kid on the block, pervasive computing applications do not have many dedicated tools at their disposal, unlike the P2P and wireless networks M&S communities. These tools range from general purpose tools such as Opnet, OMNET++ and NS 2 to customized tools for particular domains such as TOSSIM (Tiny OS SIMulator) for Wireless Sensor Networks. Designed from the grounds up for modeling computer networks, these specialized tools satisfy the basic requirements of modeling and simulation. However, the main focus of these tools is to model and simulate computer networks only. On one hand, we have complex problems where humans and mobile systems/users are involved, and on the other, we have pervasive computing and large scale global networks on an Internet-scale. Pervasive computing, being a set of immersive technologies, always has a need for tools with stronger modeling capabilities. The modeler needs the ability to work at a higher level of abstraction instead of just being able to tweak the network parameters. Tools are needed which are flexible enough to conveniently model and simulate complex interactions and protocols with ease. Such tools need to have inherent strengths and should have demonstrated abilities to model self-organization techniques, engineered emergence and other socio- and bio-inspired techniques including domains such as Ambient Intelligence [8]. Another area of application of such tools and techniques is large networks such as unstructured P2P systems which tend to grow in unpredicted dimensions. A recent example is of Gnutella where strong emergence and unexpected/unexplored graphs have been discovered quite recently[1]. Although, in these domains, tools primarily designed for modeling of computer networks can, occasionally, be used to model and simulate, but the norm is to have various levels of add-ons; this, however, is not exactly an easy marriage. The designers, being human, feel over-constrained by the limitations of having to keep tweaking and tracking low-level network parameters when their primary focus is on modeling of more abstract concepts. It is important to note that the problem emerges because these concepts are natural for the designer but alien for the tool; modern complex networks M&S specialists do not only require the ability to change sensor ranges and other basic parameters, they also need abilities to model and simulate a myriad of entities and techniques; humans, mobile robots, animals, habitat interaction, as well as self-* techniques (self-organization, self-healing, self-adaptive etc.). This could eventually be extended to any conceivable concept that could be expected to form a part of the system. So, in this domain, even if existing network oriented tools can, somehow, be used to model these concepts, the concept is typically an add-on and an after-thought to the design. Tools, which are not specifically developed to cater for complex simulations and situations, make the M&S specialist work longer to fulfill these requirements.

In contrast to this, in another part of the modeling and simulation landscape, there are a set of tools focusing on agent-based modeling and simulation techniques. These are a mature set of tools whose utility has been demonstrated in various domains. They are used primarily for M&S of complex systems' simulation models ranging from social networks, groups to self-organizing systems and bio-inspired models.

The important question we address in this paper is the following: Is there any value-addition to the use of generic and highly flexible agent-based modeling tools (which are already known to be very effective in modeling self-organization and complex systems) in the modeling and simulation of communication networks such as P2P, wireless and pervasive networks?

---

[1] http://personalpages.manchester.ac.uk/staff/m.dodge/cybergeography/atlas/topology.html

The rest of the paper is organized as follows: Section 3 briefly reviews the background and distinction of agent-based M&S tools. In Section 4, we introduce NetLogo and evaluate its usefulness using a number of modeling and simulation experiments. Finally, some concluding remarks are given in Section 5.

## 3. Review of Agent-Based Tools

### 3.1 Are there any existing Tools?

Before we start our detailed analysis on the use of NetLogo in this particular domain, it is pertinent to mention some of the previous tools in use use by network designers. Literature mentions a large set of tools, which range anywhere from customized scripts such as GnutellaSim[2] which runs on general purpose tools (such as NS-2), to Packet Level Discrete Event Simulators for simulation of similar networks as in Structella.

### 3.2 Agent-Based Tools

In the arena of agent-based tools, a number of popular tools are available. These range from Java based tools such as Mason to Logo based tools such as: StarLogo, NetLogo [3] etc. Each of these tools has different strengths and weaknesses. In the rest of the paper, we focus on just one of these tools: NetLogo as a representative of this set. We attempt to critically evaluate NetLogo and see how it fares when compared with other standard tools that are already in use for modeling and simulation of communication networks. Building on the experience of previous tools based on the Logo language, NetLogo has been developed from grounds up for complex systems research. Historically, NetLogo has been used for modeling and simulation of complex systems including multi-agent systems, social simulations, biological systems etc, on account of its ability to model problems based on human abstractions rather than purely technical aspects. However, it has not been widely used to model computer networks especially by any mainstream practitioners in the computer networks M&S domain.

### Agent-based Thinking

In a distributed environment with large scale complex interacting entities, sometimes it makes more sense to work on a simulation based on local parameters of each network node and assuming intelligence at a local level. Now, for the network modeling specialist to model a system in terms of agents, the nodes, humans or robots or any other entities which might be part of the model, need to be directly addressable. The designer of the system should be free to address one or more types or breed of objects/entities/nodes of the system. This kind of addressability is the norm in agent-based modeling and simulation.

As an example, it is fairly easy to address one or more types of entities (called agents or turtles in Logo-speak) just by their name. In other words, programs in NetLogo[3] are produced at a higher level of abstraction. Instead of the regular paradigms of looping through objects and executing functions, the designer can ask entities/nodes to move or change colors, or create links with other entities without worrying about the low-level details of how the animation or actual interaction is going to be executed. This paradigm actually produces small yet functional programs. An important benefit of small programs is their ability to greatly reduce the tweak-test-analyze cycle. This age-old paradigm of development (coming from the heritage of interpreted languages) where paper-based models are assisted by quick re-modeling of the model to fit the given parameters for verification and validation is very obvious in NetLogo. Thus it is more likely to model complex paradigms within a short time without worrying about the lower layers of other parameters unless they are important for the particular application. As we shall see in this article, it also has great expressive power and most of all, is fun to use. Although being used very frequently to model complex and self-organizing systems it has not previously been used extensively to model computer networks.

---

[2] http://www.cc.gatech.edu/computing/compass/gnutella/usage.html

[3] NetLogo is based on the Logo language

## Distinction of Agent Based M&S tools

In this section, we distinguish the power of agent-based techniques as compared to previous tools in general and NetLogo in particular. There are certain metrics by which we can evaluate Agent Based Tools and compare them with their counter parts in the domain of computer networks M&S. These metrics, however, are of a qualitative nature. The existing tools for network based M&S are not specifically tailored to developing self-organizing or complex system simulations. The M&S tools from the agent based M&S domain have several strong features such as direct addressability of nodes, ease of implementation and evaluation of self-organization, emergence and bio-inspired algorithms etc. as well as the capability of being understandable from the human perspective (having their background rooted in social simulation), all of which make them extremely useful for application in the domain of ad-hoc, P2P and pervasive systems. One way of looking at this is that originally, the object oriented programming features of the "C++" programming language were developed entirely in C in the form of a front-end which would compile to C code (as an academic exercise), however it was too cumbersome and counter-productive for programmers to follow the same practice in their regular development. It was unfeasible for companies to develop programs with object-oriented features using pure C in general. However, with the advent of modern languages with in-built features of object-orientation, it has become much easier and "natural" for software engineers to start with a template based on an object-oriented paradigm. Using the same analogy, self-organization and emergence techniques are even more abstract than object-orientation but they do entice and fit well with the human nature[4]. So, the significance of agent-based tools is to acquire the level of comfort of the M&S specialist in designing complex paradigms such as self-organization.

## Complex interaction protocols modeling.

Another point to note here is that in NetLogo, modeling complex protocols does not have to be limited to the simulation of networks alone; it can readily be used to model human users, intelligent agents, mobile robots interacting with the system or virtually any concept that the M&S designer feels worthwhile having in the model. NetLogo, in particular, has the advantage of LISP like list-processing features (with LISP being one of the most commonly used languages in the AI literature). Thus modeled entities can be shown to be interacting with the computer networks all at the same time and with the same ease that is there for modeling networks. Alternatively, the simulationist can interact and create run-time agents to interact with the network to experiment with complex protocols, which are not otherwise straightforward to conceive in terms of programs.

As an example, let us suppose, if we were to model the number of human network managers (e.g. from 10 to 100) attempting to manage a network of 10000 nodes by working on workgroups the size of n nodes (e.g. ranging from 5 to 100) at one time while giving a total of 8 hour shifts with network attacks coming in as a Poisson distribution; this could all probably be modeled in less than a few hours with only a little experience in NetLogo per se. The simulation can then be used to evaluate policies of shifts to minimize network attacks.

Another example could be the modeling and simulation of link backup policies in case of communication link failures in a complex network of 10,000 nodes along with network management policies based on part-time network managers carrying mobile phones for notification and management vs. full-time network managers working in shifts etc. all in one simulation model. And to really make things interesting, we could try these out in reality by connecting the NetLogo model to an actual J2ME based application in phones using a Java extension; so the J2ME device sends updates using GPRS to a web server which is polled by the NetLogo program to get updates while the simulation is updated in a user interface provided by NetLogo. Again, although the same could be done by a team of developers in a man-year or so of effort using different technologies, NetLogo provides for coding these almost right out of the box and the learning curve is not steep either.

---

[4] As obvious by their prevalence in Social Simulation literature

This is the expressive power of NetLogo, which lies in the sense of modeling even non-network concepts such as pervasive computing where human mobile users (e.g. in the formation of ad-hoc networks for location of injured humans) or Body Area Networks come in play along with the network. Now, it is important to note here that simulation would have been incomplete without effective modeling of all related concepts which come into play. Depending upon the application, these could vary from ambulances, doctors, nurses to concepts such as laptops, injured humans etc. in addition to readily available connectivity to GIS data provided by NetLogo extensions.

### Range of input values:

Being a general purpose tool, the abstraction level of NetLogo is specifically much higher. As such, the concepts of nodes, antenna patterns and propagation modeling are all user-dependent. On one hand, this may look burdensome to the user accustomed to using these on a regular basis, as it might appear that he or she will be working a little extra to code these in NetLogo modeling. On the other hand, NetLogo allows for the creation of completely new paradigms of network modeling, wherein the M&S specialist can focus on, for example, purely self-organization aspects or on developing antenna patterns and propagation modeling - directly in NetLogo, a relatively trivial task per se.

### Range of statistics:

NetLogo is quite flexible in terms of statistics and measurements. Any variable of interest can be added as a global variable and statistics can be generated based on single or multiple-run. Plots can be automatically generated for these variables as well.

### Handling Complex metrics:

Measurements of complex terms in NetLogo programs are limited only by the imagination of the M&S specialist. Almost any concept, which can be conceived as important, can easily be added to the model. As an example, if it is required to have complex statistics such as network assembly time, global counters can be used easily for this. Similarly, statistics such as network throughput, network configuration time, throughput delay can be easily modeled by means of similar counters (which need not be integral). By default, NetLogo provides for real-time analysis. Variables or reporters (functions which return values) can be used to measure real-time parameters and the analyst can actually have an interactive session with the modeled system without modifying the code using the "Command Window".

## 4. NetLogo Tutorial

In this section, we introduce NetLogo and demonstrate its usefulness using a number of modeling and simulation experiments.

### 4.1 What is NetLogo and how to get it?

NetLogo is a popular tool based on the Logo language with a strong user base and an active publicly supported mailing list. It provides visual simulation and is freely available for download [3] and has been used considerably in multi-agent systems literature [4]. It has also been used considerably in social simulation and complex adaptive networks [5]. One thing which distinguishes NetLogo from other tools is its very strong user support community. Most times, you can get a reply from someone in the community in less than a day. The current version of NetLogo is 4.0.4; the higher number actually demonstrates its stability and active development. NetLogo also contains an enormous number of code samples and examples. Most of the time, it is rare to find a problem for which there is no similar sample freely available either within NetLogo's model library or elsewhere from the NetLogo M&S community.

## 4.2 Basic structure of a NetLogo program

### The NetLogo world

Based on the Logo language, the NetLogo world consists of an interface which is made up of "patches". The inhabitants of this world can range from turtles to links. In addition, one can have user interface elements such as buttons, sliders, monitors, plots and so on.

### The NetLogo interface

NetLogo is a visual tool and is extremely suitable for interactive simulations. When one first opens up a NetLogo screen, an interface with a black screen is visible. There are three main tabs and a box called the command center. Briefly, the interface tab is used to work on the user interface manually and the "Information" tab is used to write the documentation for the model. And finally the "procedures" tab is where the code is actually written. The "command center" is a sort of an interactive tool for working directly with the simulation. It can be used for debugging as well as trying out commands similar to the interpreter model which, if successful, can be incorporated in one's program. The commands of a NetLogo procedure can be executed in the following main contexts:

### *Turtle*

The key inhabitants of the Logo world are the turtles which, from our perspective of designing networks can be used to easily model network nodes. The concept of agents/turtles is to provide a layer of abstraction similar to the layer of abstraction which object-oriented programming adds to structured programming paradigms. In short, the simulation can address much more complex paradigms which include pervasive models, environment or terrain models or indeed any model the M&S specialist can conceive of without requiring much additional add-on modules. However, the tool is extensible and can be directly connected to Java based modules. By writing modules using Java, the tool can potentially be used as the front end of a real-time monitoring or interacting simulation. For example, we could have a java based distributed file synchronization system, which reports results to the NetLogo interface and vice versa, the NetLogo interface could be used by the user to setup the simulation at the backend (e.g. how many machines, how many files to synchronize and subsequently with the help of the simulation, the user could simply monitor the results). Although the same can be done with a lot of other tools and technologies, the difference is that NetLogo offers these facilities almost out of the box and requires minimal coding besides being non-commercial, free and easy-to-install.

### *Patch*

A single place where the turtle exists is a patch.

### *Observer*

This is a context, which can be used in general without relating to either a patch or a turtle. The NetLogo user manual, which comes pre-packaged with NetLogo, says: "Only the observer can ask all turtles or all patches. This prevents you from inadvertently having all turtles ask all turtles or all patches ask all patches, which is a common mistake to make if you're not careful about which agents will run the code you are writing."

### Explanation of NetLogo nomenclature:

Inside the NetLogo world, we have the concept of agents. Coming from the domain of complex systems, all agents inside the world can be addressed in any conceivable way, which the user can think of, e.g., if we want to change the color of all nodes with communication power less than 0.5W, a user can simply say: ask nodes with [power < 0.5] [set color green] or if a user wants to check nodes with two link neighbors only, this can be done easily too and so on. The context of each command is one of three. Observer object is the context, when the context is neither turtle nor the patch. It is called the observer because this can be used in an interactive simulation where the simulation user can interact in this or other context using the command window.

## *The setup and the go buttons*

Although there are no real rules to creating a NetLogo program, one could design a program to have a set of procedures which can be called directly from the command center. However, in most cases, it suffices to have user interface buttons to call procedures. For the sake of this article, we shall use the standard technique of buttons.

In our program, to start with, we shall have two key buttons; Setup and the Go buttons. The idea is that the "setup" button is called only once and the "go" button is to be called multiple times (automatically). These can be inserted easily by right clicking anywhere on the interface screen and selecting buttons. So, just to start with NetLogo, the user will need to insert these two buttons in his or her model, remembering to write the names of the buttons in the commands. For the go, we shall make it a forever button. A forever button is a button which calls the code behind itself repeatedly.

Now, the buttons show up with a red text. This is actually NetLogo's way of telling us that the commands here do not yet have any code associated with them. So, let us create two procedures by the name of "setup" and "go". The procedures are written, as shown in Figure 1a, in the procedures tab and the comments (which come after a semi-colon on any line in NetLogo) explain what the commands do:

1. to setup
2. ca ; Clears everything so if we call setup again, it won't make a mess
3. crt 100 ;This means we are creating a 100 turtles
4. [
5. setxy random-pxcor random-pycor   ;These 100 turtles, we want them to be spaced out at random
6. ; patch x and y co-ordinates
7. ]
8. let mycolor random 140 ; Randomly select a color value from 0 to 139
9. ask patches
10. [
11. set pcolor mycolor ; Ask all patches to set their color to this random color
12. ]
13. end

Fig 1 (a) Code for setup

The code in Figure 1a creates 100 turtles (nodes in our case) on the screen. However the shape is a peculiar triangle by default and colors are assigned at random. Notice that we have written code here to have the patches colored randomly.

To create the procedure for the go, write the code as listed in Figure 1(b).

1. to go
2. ask turtles
3. [
4. fd 0.001 ; ask each turtle to move a small step
5. ]
6. end

Fig 1 (b) Code for go

Now, if we setup and then press go, we see turtles walking slowly in a forward direction on the screen, a snapshot of which is shown in figure 2.

Fig 2: Mobile nodes

## 4.3 Modeling Self-Organization

### Experiment I: Modeling flocking of Mobile Nodes

In this model, flocking of mobile wireless nodes within a certain radius of a transmission tower is simulated. There are two types of nodes; one type is spatially fixed and corresponds to the transmission towers. The second type of nodes perform a random walk until they come within a certain (pre-specified) radius of one of the towers, when they stop moving and change their color to the transmission tower's color. We are using this behavior to model flocking of wireless nodes within a certain radius of a tower. These are illustrated in Figure 3.

Fig 3: (a) Towers (flags) and nodes. (b) Nodes after flocking

## 4.4 Modeling complex paradigms in P2P systems

In the next set of experiments, we demonstrate the use of NetLogo for modeling and simulation of complex paradigms in P2P systems.

### Experiment II: Clustering of nodes in a P2P system

Clustering is an important technique extensively used in the literature [6]. In this example, we demonstrate the use of NetLogo for clustering in P2P systems. To generate this self-organizing behavior, nodes initiate messaging between nodes. The nodes first discover other nodes within their sensing radius. Next, nodes perform an election algorithm looking for a node with the lowest ID (as is the norm in election algorithms). Figures 4(a) and (b) illustrate respectively, the initial unclustered nodes and clustered nodes after applying self-organization.

Fig 4 (a) Initial unclustered Nodes. (b) Clustered nodes after applying self-organization

### Experiment III: Modeling Unstructured Overlay networks

Here we show how NetLogo can be used to model real-world examples of unstructured overlay forming algorithms such as Gnutella, random walk in P2P networks as well as pervasive environments as shown in [7]. Figure 5(a) shows the expanding ring algorithm being applied to a lattice. Expanding ring algorithm is an advanced form of flooding algorithm where flooding is first performed with a TTL value of 1 and then subsequently 3, 5 and so on, till one of the queries reaches the destination. Figure 5b shows the application of an advanced form of random walk, namely the k-Random walk with check, where there are k random walkers. The idea is that every few hops, these random walkers (queries) send a message back to the source node to verify if the query needs to continue its quest or not.

### Experiment IV: Distributed Averaging using gradients

In this final experiment, we first show, in Figure 6a, how NetLogo can be used to evaluate distributed consensus formation using gradients. Next, we illustrate gradient formation in wireless sensors Fig 6b. Fast self-healing gradients are discussed in [10]

Figs 5 (a) Expanding ring algorithm in a lattice. (b) k-Random walk with check on a randomly connected graph.

Fig 6 (a) Mobility based consensus formation in n=1000 nodes. (b) Gradients formed around wireless sensors

## 5. Conclusions

In this paper, we have demonstrated the use of NetLogo, an increasingly popular tool for agent based modeling in the M&S communities, from the perspective of modeling and simulation of self-organizing mobile and P2P systems. We have illustrated the strengths and distinctive aspects of agent based modeling tools by employing a number of simple experiments to demonstrate the utility of the NetLogo tool in the important domain of modeling and simulation of P2P and ad-hoc networks. In summary, we conclude that NetLogo is not only extremely flexible with a very short learning curve, but its powerful generic capabilities can be readily exploited to model self-organization in any complex system.

### Acknowledgements

We would like to thank Prof. Jose Vidal for his suggestion to try out NetLogo and Prof. Michael Huhns for helping in selection of agent-based toolkits, both from the University of South Carolina, USA.

### Author Biographies

Muaz Niazi obtained his BS in Electrical Engineering from NWFP UET, Pakistan and a Masters degree in Computer Science from Boston University, USA in 1996 and 2004 respectively. After working in industry for several years, he is currently an Assistant Professor of Software Engineering at Foundation University, Pakistan as well as being a Doctoral student at the University of Stirling, Scotland, UK. He is a member of the IEEE Communications Society and the IEEE Computational Intelligence Society. He is also a member of the IEEE CIS Task Force on Intelligent Agents and the IEEE CIS Task Force on Organic Computing. He has served on the program committee of a number of conferences and workshops such as BADS (Bio-Inspired Algorithms for Distributed Systems), IEEE Wireless Communications and Networking Conference (WCNC 2009) as well as the IEEE CIS Symposium on Intelligent Agent where he is also co-Chair of a special session on self-adaptive agents. He is also a reviewer for international Journals including the Elsevier Journal of Network and Computer Applications. His areas of research interest are modeling and simulation of self-organizing and self-adaptive systems in the P2P and ad-hoc networks domain, and novel applications of socially-inspired techniques.

Amir Hussain obtained his BEng (with 1st Class Honours) and PhD, both in Electronic and Electrical Engineering from the University of Strathclyde in Glasgow, Scotland, in 1992 and 1996 respectively. He is currently a Reader in Computing Science at the University of Stirling in Scotland. His research interests are mainly inter-disciplinary and include machine learning and cognitive computing for modeling and control of complex systems. He holds one international patent in neural computing, is Editor of five Books and has published over 100 papers in journals and refereed international Conference proceedings. He is a Senior Member of the IEEE, Editor-in-Chief of Springer's Cognitive Computation journal, Associate Editor for the IEEE Transactions on Neural Networks and serves on the Editorial Board of a number of other journals. He is IEEE Chapter Chair of the UK & RI IEEE Industry Applications Society and is a Fellow of the UK Higher Education Academy.

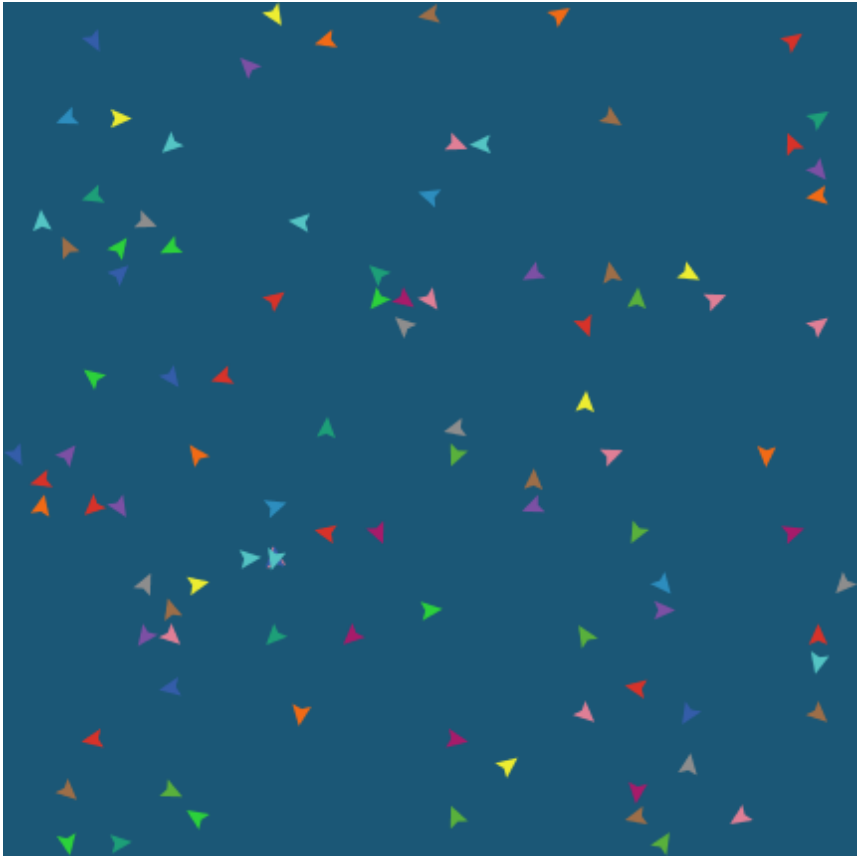
Fig 2

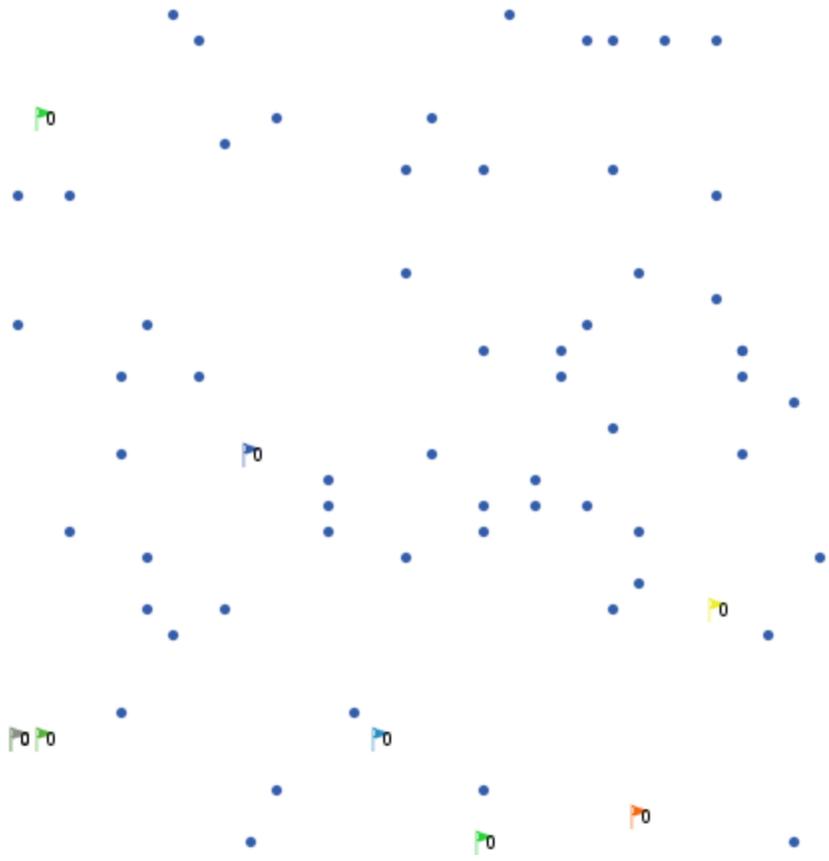

Fig 3a

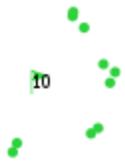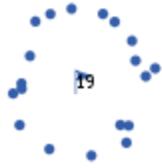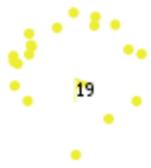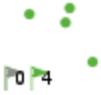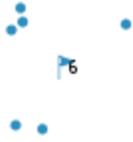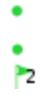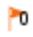

Fig 3b

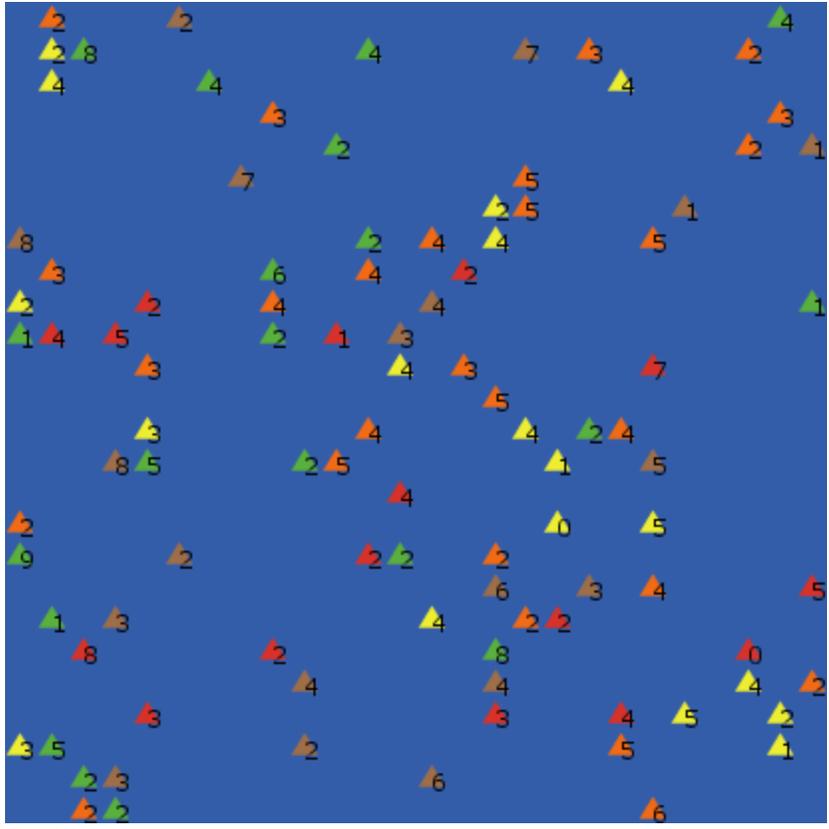

Fig 4a

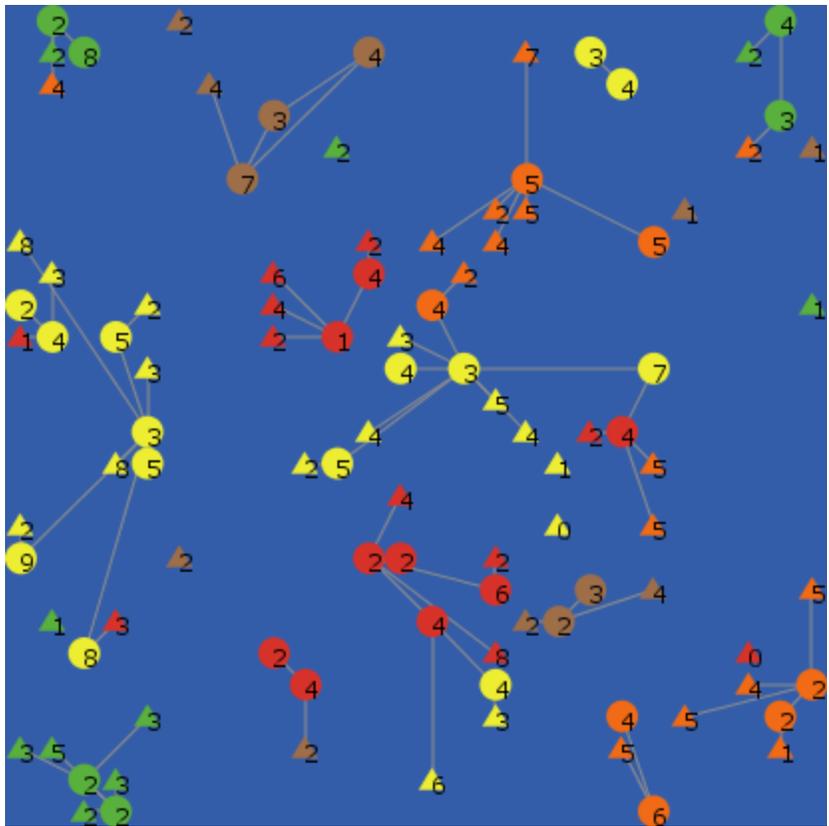
Fig 4b

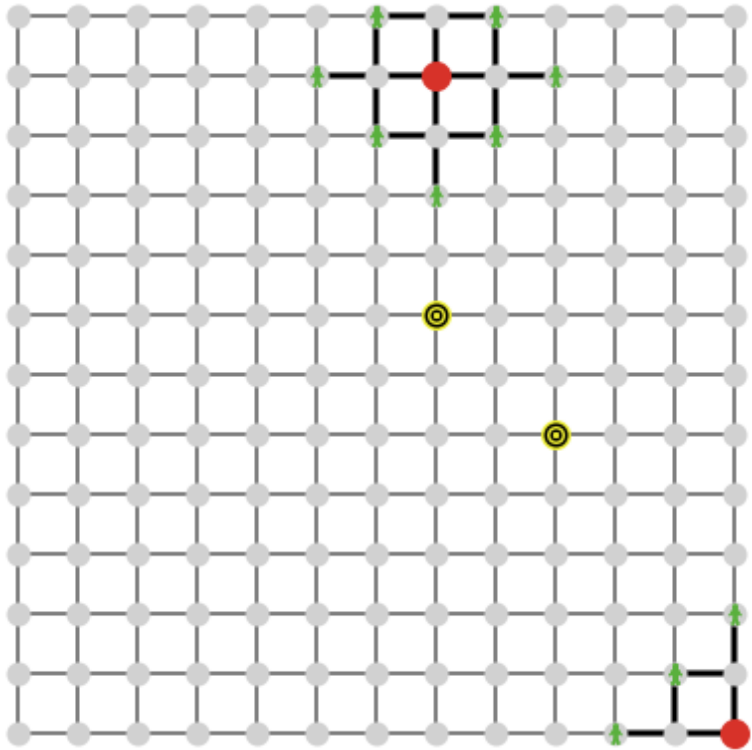

Fig 5a

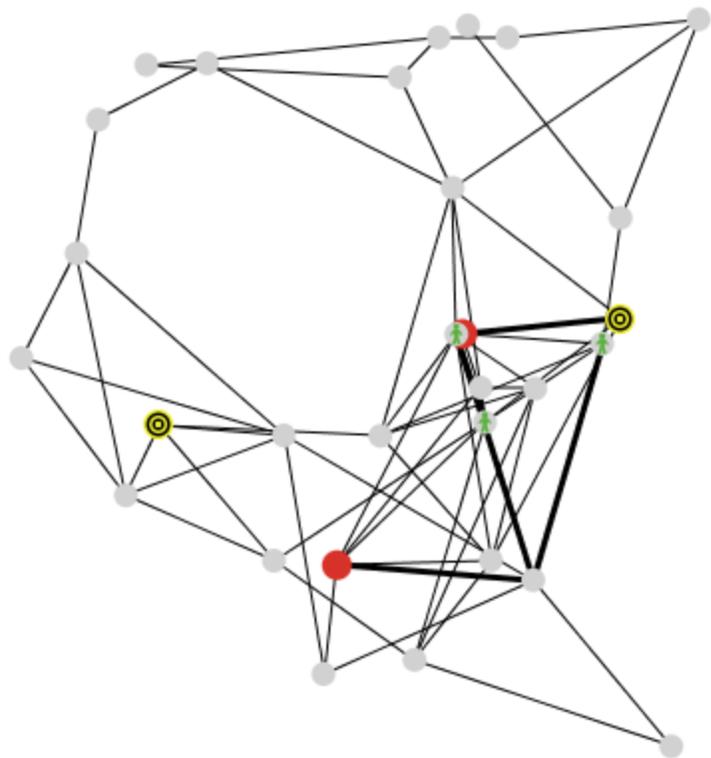

Fig 5b

Fig 6a

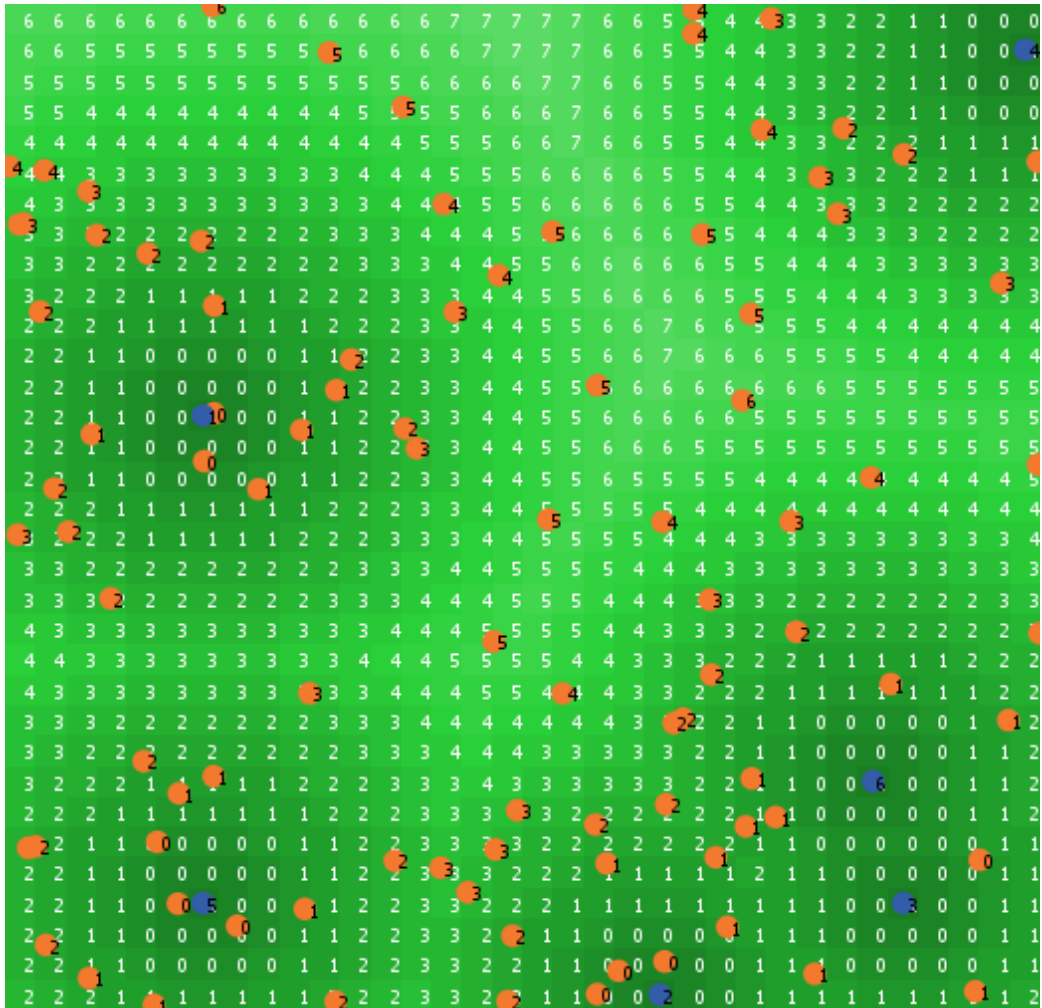

Fig 6b